\documentclass[article,aps,nofootinbib,showpacs,amsfonts,epsf]{revtex4}

\input{epsf.tex}

\newcommand{\E}{{\cal{E}}}

\renewcommand{\d}{{\rm d}}
\renewcommand{\a}{\alpha}

\newcommand{\be}{\begin{equation}}
\newcommand{\ee}{\end{equation}}
\newcommand{\bea}{\begin{eqnarray}}
\newcommand{\eea}{\end{eqnarray}}
\newcommand{\ba}{\begin{array}}
\newcommand{\ea}{\end{array}}
\def\J#1#2#3#4{{#1} {\bf #2}, #3 (#4)}

\def\PR{Phys. Rev.}
\def\PRL{Phys. Rev. Lett.}

\def\JMP{J. Math. Phys.}

\def\CQG{Class. Quantum Grav.}

\begin{document}
\draft
\title{Formation of a Kerr black hole from two stringy NUT objects}

\author{V.~S.~Manko,$^*$ E.~D.~Rodchenko,$^\dagger$ E.~Ruiz$\,^\ddagger$ and M.~B.~Sadovnikova$\,^\dagger$}
\address{$^*$Departamento de F\'\i sica, Centro de Investigaci\'on y de
Estudios Avanzados del IPN, A.P. 14-740, 07000 M\'exico D.F.,
Mexico\\$^\dagger$Department of Quantum Statistics and Field
Theory, Lomonosov Moscow State University, Moscow 119899,
Russia\\$^\ddagger$Instituto Universitario de F\'{\i}sica
Fundamental y Matem\'aticas, Universidad de Salamanca, 37008
Salamanca, Spain}

\begin{abstract}
In this paper we show that an isolated Kerr black hole can be
interpreted as arising from a pair of identical counter--rotating
NUT objects placed on the symmetry axis at an appropriate distance
from each other. \end{abstract}


\maketitle

As has been recently shown \cite{MRu}, the well--known NUT
solution \cite{NTU} represents a stringy object with two
counter--rotating semi--infinite {\it massive} singularities
attached to the poles of a central non--rotating mass. This
solution is the simplest possible one among the equatorially
antisymmetric spacetimes the notion of which has been introduced
in \cite{EMR}. Although the usual NUT object is endowed with some
unphysical properties, e.g., the presence of the regions with
negative mass and closed time--like curves, in the present paper
we will demonstrate that configurations of several NUT
constituents can give rise to physically significant models
without the pathologies of a single NUT spacetime. Concretely, we
shall consider a simple but convincing example: emergence of a
Kerr black hole from two interacting NUT objects. The
consideration will be carried out within the framework of an exact
solution describing a non--linear superposition of two NUT
sources.

We remind that the stationary axisymmetric vacuum problem reduces
to finding the complex function $\E$ satisfying the Ernst equation
\cite{Ern} \be
(\E+\bar\E)(\E_{,\rho,\rho}+\rho^{-1}\E_{,\rho}+\E_{,z,z})=
2(\E_{,\rho}^2+\E_{,z}^2), \label{E_eq} \ee where $\rho$ and $z$
are the Weyl--Papapetrou cylindrical coordinates, a comma denotes
partial differentiation with respect to the coordinate that
follows it, and a bar over a symbol means complex conjugation.
Using Sibgatullin's method \cite{Sib}, the potential $\E$
satisfying (\ref{E_eq}) can be constructed from its value on the
upper part of the symmetry axis. For $z>\sqrt{m^2+\nu^2}$, the
Ernst complex potential of a single NUT solution has the form
\cite{MRu} \be e(z)\equiv\E(\rho=0,z)= \frac{z-m-i\nu}{z+m+i\nu},
\label{nut_axis} \ee where $m$ is the total mass of the stringy
NUT object, while $\nu$ represents the average angular momentum
per unit length of the semi--infinite NUT singularity. The
non--linear superposition of two NUT solutions with equal masses
and opposite angular momenta is formally defined by the axis data
\be e(z)=
\frac{z-k-m-i\nu}{z-k+m+i\nu}\cdot\frac{z+k-m+i\nu}{z+k+m-i\nu},
\label{nut2_axis} \ee the real parameter $k$ representing a
displacement of each NUT constituent along the $z$--axis. We point
out that whereas the axis data (\ref{nut_axis}) define the
equatorially antisymmetric solution because of the characteristic
relation $e(z)e(-z)=1$ fulfilled in this case \cite{EMR}, the axis
data (\ref{nut2_axis}) satisfy the relation $e(z)\bar e(-z)=1$ and
hence define already the {\it equatorially symmetric} spacetime
\cite{Kor,MNe}.

Below we give the final form of the Ernst potential $\E$
constructed from the data (\ref{nut2_axis}) with the aid of
Sibgatullin's method, and the form of all the corresponding metric
functions $f$, $\gamma$ and $\omega$ from the Papapetrou
stationary axisymmetric line element \be \d
s^2=f^{-1}[e^{2\gamma}(\d\rho^2+\d z^2)+\rho^2\d\varphi^2]-f(\d
t-\omega\d\varphi)^2, \label{papa} \ee omitting the intermediate
steps (we refer the reader to the paper \cite{MRu1} for details):
\bea \E&=&\frac{A-B}{A+B}, \quad f=\frac{A\bar A-B\bar
B}{(A+B)(\bar A+\bar B)}, \quad e^{2\gamma}=\frac{A\bar A-B\bar
B}{64d^4\a_+^2\a_-^2R_+R_-r_+r_-}, \quad \omega=-\frac{4{\rm
Im}[G(\bar A+\bar B)]}{A\bar A-B\bar B},
\nonumber\\
A&=&[(m^2+\nu^2)(k^2-m^2)(k^2-m^2-\nu^2)-2m^2k^2\nu^2](R_+-R_-)(r_+-r_-)
\nonumber\\ &+&2\a_+\a_-(m^2+\nu^2)(m^2-k^2)(R_+R_-+r_+r_-)
\nonumber\\
&-&\a_+\a_-[2m^4+(m^2+\nu^2)(k^2-m^2)](R_++R_-)(r_++r_-)
\nonumber\\ &-&2imk\nu
d[(\a_+-\a_-)(R_+r_+-R_-r_-)-(\a_++\a_-)(R_+r_--R_-r_+)],
\nonumber\\ B&=&4d\{ m\a_+\a_-[(m^2-d)(R_++R_-)-(m^2+d)(r_++r_-)]
\nonumber\\ &+&ik\nu[\a_-(m^2-d)(R_+-R_-)-\a_+(m^2+d)(r_+-r_-)]\},
\nonumber\\
G&=&d[d^2+m^2(m^2+2ik\nu)][(\a_+-\a_-)(R_-r_--R_+r_+)+(\a_++\a_-)(R_+r_--R_-r_+)]
\nonumber\\
&-&2m^2d^2[(\a_++\a_-)(R_-r_--R_+r_+)+(\a_+-\a_-)(R_+r_--R_-r_+)]
\nonumber\\ &-&m\a_+\a_-(d^2+m^4)(R_++R_-)(r_++r_-)
+m[kd^2(k+4i\nu)-(2k^2-m^2)(m^2+\nu^2)^2 \nonumber\\
&+&k^2\nu^4](R_+-R_-)(r_+-r_-)-2m\a_+\a_-(k^2-m^2)(m^2+\nu^2)(R_+R_-+r_+r_-)
\nonumber\\ &-&2dz\{\a_-(m^2-d)[m\a_+(R_++R_-)+ik\nu(R_+-R_-)]
\nonumber\\ &-&\a_+(m^2+d)[m\a_-(r_++r_-)+ik\nu(r_+-r_-)]\}
\nonumber\\
&+&2d\a_+\a_-(2m^2+ik\nu)[m^2(R_++R_--r_+-r_-)-d(R_++R_-+r_++r_-)]
\nonumber\\
&+&2md[d^2-m^4-ik\nu(2m^2+ik\nu)][\a_-(R_--R_+)+\a_+(r_+-r_-)]
\nonumber\\
&-&2md^2(m^2-k^2+\nu^2-2ik\nu)[\a_-(R_--R_+)+\a_+(r_--r_+)],
\label{metric}  \eea where \be R_\pm=\sqrt{\rho^2+(z\pm\a_+)^2},
\quad r_\pm=\sqrt{\rho^2+(z\pm\a_-)^2}, \label{ri} \ee and \be
\a_\pm=\sqrt{m^2+k^2-\nu^2\pm 2d}, \quad
d=\sqrt{m^2k^2+\nu^2(k^2-m^2)}. \label{d} \ee

The metric obtained is asymptotically flat and it describes a
spinning body with the total mass $2m$, total angular momentum
$2k\nu$ and mass--quadrupole moment $2m(k^2-m^2-\nu^2)$. Since the
well--known Kerr solution \cite{Ker} is characterized by the
parameters $m$ and $a$ (total mass and angular momentum per unit
mass, respectively, the mass--quadrupole moment of the source
being equal to $-ma^2$), one could suppose that in the particular
case $k=m$ the metric (\ref{metric}) reduces to the Kerr solution
with the total mass $2m$ and total angular momentum per unit mass
$\nu$. To see whether this supposition is true, let us put $k=m$
in (\ref{metric}) and (\ref{d}). Then we get \be
\a_\pm=\sqrt{2m^2-\nu^2\pm 2m^2}, \quad d=m^2, \ee and the
expressions for $A$, $B$ and $G$ take the form \bea
A&=&[(\a_+-i\nu)R_++(\a_++i\nu)R_-]F, \nonumber\\ B&=&4m\a_+F,
\nonumber\\
G&=&m[(\a_+-2m-i\nu)R_++(\a_++2m+i\nu)R_-+2\a_+(2m+i\nu -z)]F,
\nonumber\\ F&=&-2m^4[(\a_-+i\nu)r_++(\a_--i\nu)r_-], \eea $F$
being the common factor. Cancelling out $F$ in the Ernst potential
and in the metric functions, we arrive, after the introduction of
spheroidal coordinates \be x=\frac{1}{2\a_+}(R_++R_-), \quad
y=\frac{1}{2\a_+}(R_+-R_-), \ee at the formulas \bea
\E&=&\frac{\a_+x-i\nu y-2m}{\a_+x-i\nu y+2m}, \quad
f=\frac{\a_+^2x^2+\nu^2y^2-4m^2} {(\a_+x+2m)^2+\nu^2y^2},
\nonumber\\ e^{2\gamma}&=& \frac{\a_+^2x^2+\nu^2y^2-4m^2}
{\a_+^2(x^2-y^2)}, \quad \omega=-\frac{4m\nu(\a_+x+2m)(1-y^2)}
{\a_+^2x^2+\nu^2y^2-4m^2}. \label{metric1} \eea Introducing now
the constants \be p=\frac{\a_+}{2m}, \quad q=\frac{\nu}{2m}, \quad
p^2+q^2=1, \ee we rewrite (\ref{metric1}) in the form \bea
\E&=&\frac{px-iq y-1}{px-iq y+1}, \quad f=\frac{p^2x^2+q^2y^2-1}
{(px+1)^2+q^2y^2}, \nonumber\\ e^{2\gamma}&=&
\frac{p^2x^2+q^2y^2-1} {p^2(x^2-y^2)}, \quad
\omega=-\frac{2\a_+q(px+1)(1-y^2)}{p(p^2x^2+q^2y^2-1)},
\label{metric2}  \eea which is one of the standard representations
of the Kerr solution \cite{Ern,KSt}.

Therefore, we have demonstrated that a specific non--linear
superposition of two NUT solutions can give rise to the Kerr
spacetime. We hope that a future investigation will be able to
reveal that this result is not simply a mathematical curiosity but
could have a solid physical nature.

\vspace{1cm} This work was partially supported by Project 45946--F
from CONACyT, Mexico, by Project FIS2006--05319 from MCyT, Spain,
and by Project SA010C05 from Junta de Castilla y Le\'on, Spain.

\end{document}